\documentclass[pra,aps,groupaddress,showpacs,twocolumn]{revtex4}
\usepackage{epsfig,amsmath,graphicx,amssymb,overpic}
\usepackage{bm}
\usepackage{mathrsfs}
\usepackage{graphicx}
\usepackage{epsfig}
\usepackage{dcolumn}% Align table columns on decimal point
\usepackage{bm}% bold math
\usepackage{amsmath,amssymb,amsthm}
\usepackage[colorlinks=true,linkcolor=blue]{hyperref}
\usepackage{subfigure}
\usepackage{booktabs}
\usepackage[mathscr]{euscript}
\usepackage{ulem}

\newcommand{\bea}{\begin{eqnarray}}
\newcommand{\eea}{\end{eqnarray}}
\newcommand{\beq}{\begin{equation}}
\newcommand{\eeq}{\end{equation}}

\def\/{\over}

\begin{document}

\title{Soliton molecules and asymmetric solitons\\ in fluid systems via velocity resonance}
\author{S. Y. Lou}
\affiliation{\footnotesize{School of Physical Science and Technology, Ningbo University, Ningbo, 315211, China}}

\begin{abstract}
Soliton molecules may be formed in some possible mechanisms in both theoretical and experimental aspects. In this letter, we introduce a new possible mechanism, the velocity resonant, to form soliton molecules. Under the resonant mechanism, two solitons may be formed to a kink-antikink molecule, an asymmetric soliton, a two-peak soliton and/or a far away bounded molecule depended on the selections of the wave numbers and the distance between two solitons of the molecule.  The results are exhibited via three well known fifth order integrable systems which approximately solve a general fluid model that also appeared in many other physical fields. The interactions among multiple soliton molecules for these three integrable systems are elastic.
\end{abstract}

\maketitle

Soliton molecules, bound states of solitons, have been experimentally observed in optics \cite{PRL2005,Sci2017,LiuXM2018} and numerically predicted in Bose-Einstein condensates \cite{PRA2012}.
On the other hand, solitons play a very important role in various modern scientific fields including fluids \cite{fluid}, plasmas \cite{plasma}, fibers \cite{fiber}, optics \cite{optic}, complex networks \cite{net}, quantum field theory \cite{field}, gravity \cite{gravity}, Bose-Einstein condensates\cite{BEC}, atmospheric and oceanic dynamics\cite{LFH} and so on. In addition to optical systems, fluid systems (such as the stratified fluids \cite{HHC1979}, magnetic fluids \cite{PRL2005a}, quantum fluids \cite{PRL2011}, fluids of light \cite{RMP2013}, atomic and molecular gases \cite{RMP2017}, degenerate electron fluids \cite{RMP2011} and superfluid
$^3$He \cite{RMP1987}) exhibit abundant soliton structures and dynamics. 
Solitary waves and solitons are firstly discovered in fluid. We believe that soliton molecules should be observed in fluids such as oceanic and atmospheric systems. 

Resonance is also an important natural phenomena which may lead to a disaster/catastropere (blow up in mathematics). For integrable systems, because of the complexities introduced by nonlinearity, resonances of solitons may lead to various types of new excitations such as the breathers (analytic) \cite{breath} or complexitons (singular) \cite{Ma} (caused by the module resonance of wave numbers, say, $|k_2|=|k_1|$ i.e. $k_2=\pm k_1^*$), soliton fissions, soliton fusions \cite{WLT}, instantons/rogue waves \cite{Liu} and rational-exponential waves (caused by wave number resonance companied by vanishing procedure, say, $k_2\rightarrow \pm k_1 \rightarrow 0$) \cite{ZhangDJ}, web solitons and lumps (by wave number resonance for $\vec{k}_2=\pm \vec{k}_1$ in high dimensions) \cite{Kodama1}.
In this letter, we try to find a new mechanism to form soliton molecules by introducing velocity resonance $\omega_1/k_1=\omega_2/k_2$.

One of significant fluid models to describe surface and internal waves can be written as
\begin{eqnarray}
&&u_t+\left[\alpha u^2+\beta u_{xx}+\epsilon\big(\alpha_1 u^3 +\gamma_1 u u_{xx}+\gamma_2 u_x^2\right.\nonumber\\
&&\left.+\beta_1 u_{xxxx}\big)\right]_x+o(\epsilon^2)=0 \label{5th}
\end{eqnarray}
which is derived by many authors\cite{Lamb,GPP,KRI}. Some authors (say, Kodama\cite{Kodama} and Fokas and Liu \cite{Fokas}) have proved that the model \eqref{5th} is asymptotic integrable up to the same order $\epsilon^2$ because model \eqref{5th} can be asymptotically changed to either usual Korteweg de-Vries (KdV) equation ($\epsilon =0$) or the fifth order integrable KdV equation (the summation of the usual KdV and its next flow) by using suitable transformations. In fact, similar to the Kodama's idea \cite{Kodama}, the transformation
\begin{eqnarray}
u&=&v+\epsilon \left[\left(\frac{5a}{3 \beta^2}\alpha\beta_1 +a_1\right)v^2+\left(\frac{5b\beta_1}{2\beta}+b_1\right)v_{xx}
\right.\nonumber\\
 &&\left.+\left(\frac{5a\alpha\beta_1}{3\beta^2} -\frac{\gamma_1}{3\beta}\right)v_x\int v\mbox{d}x\right]\label{tr1}
\end{eqnarray}
with
\begin{eqnarray}
&&v_t+\left[\alpha v^2+\beta v_{xx}+\epsilon\beta_1 \left(v_{xxxx}+\frac{5a\alpha}{\beta} vv_{xx}\right.\right.\nonumber\\
&&\left.\left.+\frac{5(a-b)\alpha}{\beta} v_x^2+\frac{5a\alpha^2}{3\beta^2} v^3\right) \right]_x=0\label{gKdV}
\end{eqnarray}
where $a_1\equiv \frac{\gamma_1}{6\beta}-\frac{3 \alpha_1}{2\alpha}$, $b_1\equiv \frac{\gamma_1}{4\alpha} +\frac{\gamma_2}{2\alpha} -\frac{9\beta\alpha_1}{4\alpha^2}$ and $a$ and $b$ are arbitrary constants,
approximately solves the original equation \eqref{5th} up to the same order $\epsilon$.
To see the integrability of \eqref{gKdV}, one can make a Galileo transformation $v\rightarrow -\beta^2/(5a\alpha\beta_1\epsilon) +v(x+\beta^2t/(5a\beta_1\epsilon), t)$ such that the terms $\alpha v^2$ and $\beta v_{xx}$ in \eqref{gKdV} are vanished.

 It is well known that there are three integrable cases, the usual fifth order KdV equation for the selections $\{a,b\}_{KdV}=\left\{\frac23,\frac13\right\}$, the Sawada-Kotera (SK) model by taking $\{a,b\}_{SK}=\{1,1\}$ and the Kaup-Kupershmidt (KK) equation with $\{a,\ b\}_{KK}=\left\{1,\frac14\right\}$. The approximate transformation \eqref{tr1} for the fifth order KdV case, $\{a,b\}_{KdV}=\left\{\frac23,\frac13\right\}$, is firstly given by Kodama\cite{Kodama} while other two cases, the SK and KK cases have not yet been found elsewhere in our knowledge.

For the fifth order KdV and SK cases, the multiple soliton solutions possess the form ($\xi_j=k_jx-\omega_jt
+\xi_{j0},\ \omega_j=\epsilon\beta_1k_j^5+\beta k_j^3,\ v\rightarrow u $),
\begin{equation}
u=\frac{6\beta}{\alpha}\left\{\ln\left[\sum_{\nu}K_{\nu}
\cosh\left(\sum_{i=1}^N\frac{\nu_j\xi_{j}}2\right)
\right]\right\}_{xx},\label{NS}
\end{equation}
where the summation $\nu$ should be done for all possible permutations $\nu_j=1,\ -1,\ j=1,\ 2,\ \ldots,\ N$,
$
K_{\nu}=\prod_{i<j}a_{ij},\
$
$a_{ij}^2=(\nu_ik_i-\nu_jk_j)^2$ for the fifth order KdV case,
$a_{ij}^2=(\nu_ik_i-\nu_jk_j)^2 \big[5\epsilon\beta_1(k_i^2-\nu_i\nu_jk_ik_j+k_j^2)+3\beta\big]$
for the SK case, $k_j$ and $\xi_{j0},\ j=1,\ 2,\ \ldots,\ N$ are arbitrary constants.

Now, we are interested in finding something new from \eqref{NS} except for those from the usual KdV equation ($\epsilon=0$ in \eqref{5th}). In additional to the N soliton solutions, the expression \eqref{NS} includes many kinds of resonant excitations. All the known resonant solutions of \eqref{NS} are singular solutions with blow up properties including the complexiton solutions \cite{Ma} (or namely singular breathers caused by the resonant condition $|k_i|=|k_j|$, i.e., $k_i=\pm k_j^*$), the rational-exponential function mixed solutions \cite{ZhangDJ} (introduced by the resonant conditions $k_i=\pm k_j$) and the rational solutions (happened with further resonant conditions $k_i\rightarrow k_j \rightarrow 0$).

To find nonsingular analytical resonant excitations from \eqref{NS} we introduce a novel type of resonant conditions ($k_i\neq \pm k_j$), the velocity resonance,
\begin{equation}
\frac{k_i}{k_j}=\frac{\epsilon\beta_1k_i^5+\beta k_i^3} {\epsilon\beta_1k_j^5+\beta k_j^3}. \label{mol}
\end{equation}
Because of the resonant condition \eqref{mol} the $i^{th}$ and $j^{th}$ solitons are bounded to form a soliton molecule or an asymmetric soliton depending on the selections on the solution parameters. To see this fact, we take $N=2$ as a simple example. For $N=2$, the expression \eqref{NS} can be simplified to
\begin{equation}
u=\frac{3\beta a_- a_+ [k_2^2\cosh(\xi_1)+k_1^2\cosh(\xi_2)+a_- a_+ ]} {\alpha \left[a_-\cosh\left(\frac{\xi_1+\xi_2}2\right)+a_+ \cosh\left(\frac{\xi_1-\xi_2}2\right)\right]^2}, \label{2sol}
\end{equation}
where $\xi_i=k_ix-\omega_it+\xi_{i0},\ i=1,\ 2$, $a_{\pm}^2=(k_1\pm k_2)^2$ for the fifth order KdV case and $a_{\pm}^2=(k_1\pm k_2)^2 \big[5\epsilon\beta_1(k_1^2\pm k_1k_2+k_2^2)+3\beta\big]$ for the SK case.

Two-soliton solution \eqref{NS} exhibits one soliton molecule structure under the resonance condition \eqref{mol}. The density plot Fig. 1 displays the molecule structure expressed by \eqref{2sol} with the parameter selections ($\alpha=3,\ \beta=-\beta_1=1$ for all figures of this paper),
\begin{equation}
 k_1=2\sqrt{21}, k_2=4, \epsilon=0.01, \xi_{10}=0, \xi_{20}=20, \label{skm1}
\end{equation}
for the SK case.
\input epsf
\begin{figure}
\centering\epsfxsize=6.5cm\epsfysize=4cm\epsfbox{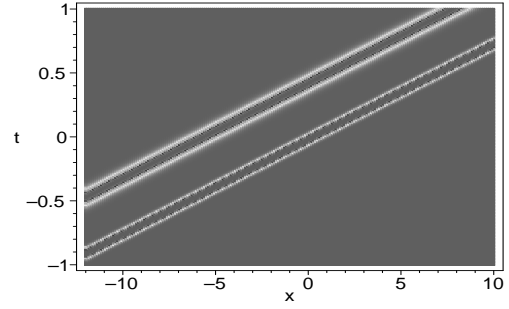}
\caption{Density plot of the soliton molecule for the SK system described by Eq. (\ref{2sol}) with the parameter
selections \eqref{skm1}.}\label{fig1}
\end{figure}
Fig. 2 is a three dimensional plot of the soliton molecule \eqref{2sol} of the fifth order KdV system with the same parameter selections \eqref{skm1}.
\input epsf
\begin{figure}
\centering\epsfxsize=6.5cm\epsfysize=4cm\epsfbox{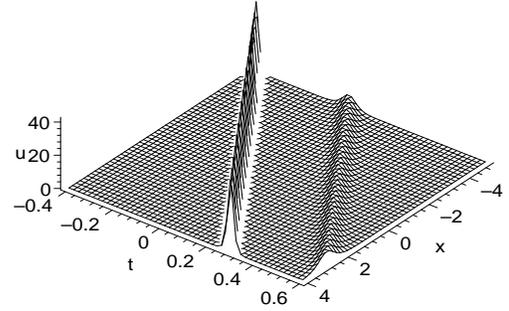}
\caption{Soliton molecule structure for the fifth order KdV system described by Eq. (\ref{2sol}) with the parameter
selections \eqref{skm1}.}\label{fig2}
\end{figure}

From Fig. 1, Fig.2 and the expression \eqref{2sol}, one can find that two solitons in the molecule are different because $k_2\neq k_1$ though the velocities of them are same.

The distance, $s$, between two solitons of the molecule depends on the parameters $\xi_{10}$, $\xi_{20}$ and $k_1$ via
\begin{equation}
s^2=\frac1{4\omega_1^{3}}\beta_1\epsilon k_1^{3}(c\xi_{10}-\xi_{20})^2
(k_1^{2}+\omega_1^2),\ \label{s}\end{equation}
where $c= \frac1{k_1^2} \sqrt{\frac{-\omega_1}{k_1\beta_1\epsilon}}$ and $ \omega_1=k_1^3(\beta+\epsilon \beta_1 k_1^2)$.

It is clear that the soliton molecule will become an asymmetric one soliton solution if the two solitons of the molecule is closed enough. Fig. 3 and Fig. 4 are the plots of the asymmetric soliton solutions related to  the fifth order KdV case and the SK case, respectively. The parameters related to Fig. 1 are selected as
\begin{equation}
k_1=2 \sqrt{21}, k_2=4, \epsilon=\frac{1}{100}, \xi_{10}=0, \xi_{20}=1.\label{pFig3}
\end{equation}
The parameters applied in Fig. 2 are same as those in Fig. 1 except for $\xi_{20}=0.1$.
\input epsf
\begin{figure}
\centering\epsfxsize=6.5cm\epsfysize=4cm\epsfbox{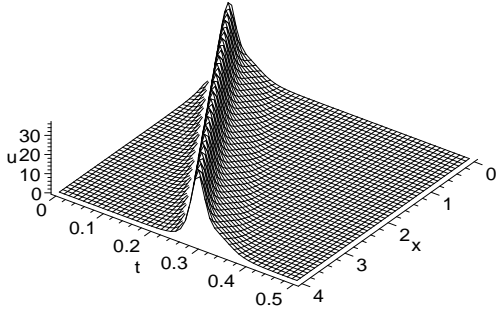}
\caption{Asymmetric soliton for the fifth order KdV equation  described by Eq. (\ref{2sol}) with the parameter
selections \eqref{pFig3}.}\label{fig3}
\end{figure}

\input epsf
\begin{figure}
\centering\epsfxsize=6.5cm\epsfysize=4cm\epsfbox{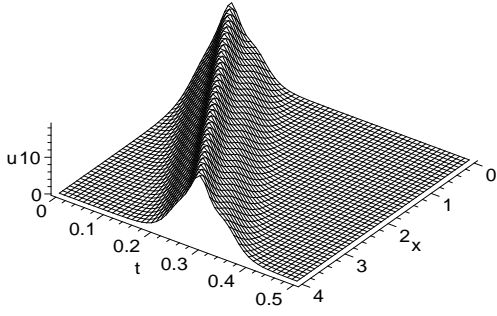}
\caption{Asymmetric soliton for the SK system described by Eq. (\ref{2sol}) with the same parameter selections as in Fig. 3 except for $\xi_{20}=0.1$.}\label{fig4}
\end{figure}

It is known that the interactions among solitons for both the SK equation and the fifth order KdV system are elastic. Thus, it is not surprised that the interactions among soliton molecules are also elastic. Fig. 5 displays the elastic interaction property for the two soliton molecule solution \eqref{NS} of the fifth order KdV equation with $N=4$ and the other parameter selections
\begin{eqnarray}
&&k_1=2 \sqrt{21},\ k_2=8,\ k_3=6,\ k_4=4,\ \epsilon=0.01,\nonumber\\
 &&\xi_{10}=\xi_{30}=0,\ \xi_{20}=20,\ \xi_{40}=15. \label{2mol}
\end{eqnarray}

\input epsf
\begin{figure}
\centering\epsfxsize=6.5cm\epsfysize=4cm\epsfbox{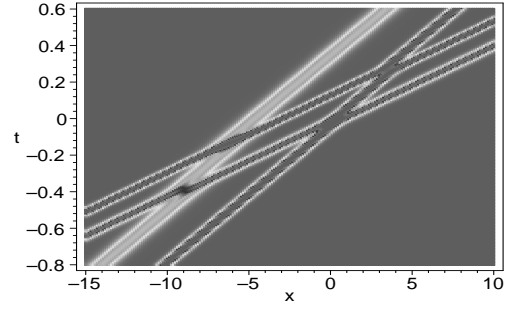}
\caption{Plot of the elastic interaction property between two soliton molecules for the fifth order KdV equation described by \eqref{NS} with $N=4,\ \alpha=3, \ \beta=\beta_1=1$ and the parameter selections \eqref{2mol}.}\label{fig5}
\end{figure}

For the KK case \eqref{gKdV} with $\{a,\ b\}=\{1,\ 1/4\}$, the multisoliton solution does not possess the form \eqref{NS} \cite{XBHu}, however, it is interesting that its resonant cases of \eqref{NS} with
\begin{equation}
N=2n,\ \frac{k_i}{k_{n+i}}=\frac{\omega_i}{\omega_{n+i}}=\pm 1,\ i=1,\ \ldots,\ n\label{rkk}
\end{equation}
do exist for all $n$. Here, we just list the details for $n=1$ and $n=2$.

The resonant solution \eqref{NS} with \eqref{rkk} and $n=1$ for the KK system \eqref{gKdV} with $\{a,\ b\}=\{1,\ 1/4\}$ possesses
the form ($v\rightarrow u,\ \xi=kx-\omega t+\xi_0$),
\begin{eqnarray}
u&=&\frac{a_0\beta}{\alpha}+\frac{a_1\beta}{\alpha} \left\{\ln\left[c+\cosh(\xi)\right]\right\}_{xx}, \label{kk2s}\\
\omega &= & 5k\beta_1\epsilon a_0^2 +k (5\beta_1\epsilon k^2+2\beta) a_0+k^3 (\beta_1\epsilon k^2+\beta),\nonumber
\end{eqnarray}
where the parameters $a_0,\ a_1$ and $ c$ can be taken in three nonequivalent ways
\begin{eqnarray}
&&a_0=-\frac{(c^2-4) k^2}{4 (c^2-1)} -\frac{\beta}{5\beta_1\epsilon},\ a_1 = 3, \label{kk2s1}\\
&&a_0 = -\frac{ k^2}{4}-\frac{\beta}{5\beta_1\epsilon},\ a_1 = \frac{3}{2},\ c^2 =1,
\label{kk2s2}
\end{eqnarray}
and
\begin{eqnarray}
&&a_0 = -2 k^2 -\frac{\beta}{5\beta_1\epsilon},\ a_1 = 12,\ c^2 =1,
 \label{kk2s3}
\end{eqnarray}
while $k,\ \xi_0$ and $c$ in the first case \eqref{kk2s1} remain free.

The resonant soliton solution \eqref{kk2s} with \eqref{kk2s1} displays abundant structures because of the existence of the free parameter $c$. For $-1<c<1$ and $1<c<2$, the solution \eqref{kk2s} with \eqref{kk2s1} can be considered as an analytic  single soliton with one peak. For $c=2$, the resonant solution \eqref{kk2s} with \eqref{kk2s1} can be considered as a bounded kink-antikink soliton (a kink-antikink molecule), or a fat soliton, and/or a plateau soliton as shown in Fig. 6 with the parameter selections
\begin{equation}
c=2,\ \epsilon=0.1,\ k=0.1,\ \xi_0=0. \label{pFig6}
\end{equation}
\input epsf
\begin{figure}
\centering\epsfxsize=6.5cm\epsfysize=4cm\epsfbox{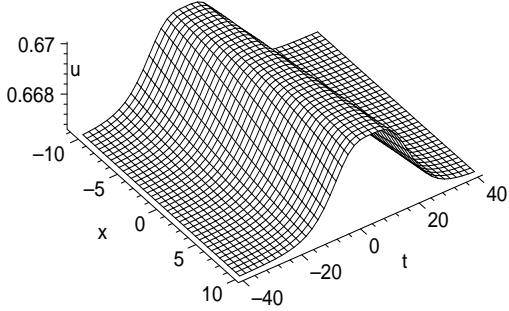}
\caption{Kink-antikink molecule for the KK equation described by \eqref{kk2s} with the parameter selections \eqref{pFig6}.}\label{fig6}
\end{figure}

This critical solution, $c=2$ case, may be called a plateau soliton or a dissipative soliton because the property
$$\left.u_x\right|_{\xi=0}=\left.u_{xx}\right|_{\xi=0}
=\left.u_{xxx}\right|_{\xi=0}=0$$
at the soliton center, $\xi=0$.

For $c>2$ the solution \eqref{kk2s} with \eqref{kk2s1} is clearly a two-soliton molecule and two peaks of the molecule will be separated as $c$ increases. Fig. 7 displays the structure of the soliton molecule \eqref{kk2s} with the parameter selections
\begin{equation}
c=3000,\ \epsilon=0.1,\ k=2.828428,\ \xi_0=0.\label{pFig7}
\end{equation}
\input epsf
\begin{figure}
\centering\epsfxsize=6.5cm\epsfysize=4cm\epsfbox{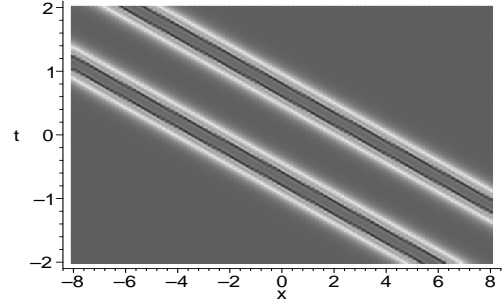}
\caption{Soliton molecule structure for the KK equation described by \eqref{kk2s} where the parameters are fixed by \eqref{pFig7}.}\label{fig7}
\end{figure}
Different from the SK and the fifth order KdV cases, two solitons in the molecule of the KK equation are completely same.

For the second and third types of soliton solutions \eqref{kk2s} with \eqref{kk2s2} or \eqref{kk2s3} are only single analytic peaked solitons for $c=1$.

The resonant solution \eqref{NS} with \eqref{rkk} and $n=2$ for the KK system \eqref{gKdV} with $\{a,\ b\}=\{1,\ 1/4\}$ possesses the form ($v\rightarrow u$),
\begin{eqnarray}
u&=&-\frac{\beta^2}{5\alpha\beta_1\epsilon}+\frac{ 3\beta}{\alpha}\left(\ln \psi \right)_{xx},\ \label{kk4s}
\end{eqnarray}
with $\psi=8(k_1^2-k_2^2)^2+12k_1^2k_2^2
+4a_{+}a_-[\cosh(\xi_1)+\cosh(\xi_2)]
+a_+^2\cosh(\xi_1-\xi_2)+a_-^2\cosh(\xi_1+\xi_2)$,
where $a_{\pm}^2=(k_1\pm k_2)^2(k_1^2\pm k_1k_2 +k_2^2)$, $\xi_i=k_ix-\left(\beta_1\epsilon k_i^5 -\frac{k_i\beta^2}{5\epsilon\beta_1})\right)t+\xi_{i0},\ i=1,\ 2$ and $k_1,\ k_2,\ \xi_{10}$ and $\xi_{20}$ are arbitrary constants.

Fig. 8 displays the interaction between two kink-antikink molecules described by \eqref{kk4s} with the parameter selections
\begin{equation}
k_1=2,\ k_2=1.6,\ \epsilon=0.1,\ \xi_{10}=\xi_{20}=0. \label{kk4s1}
\end{equation}
\input epsf
\begin{figure}
\centering\epsfxsize=6.5cm\epsfysize=4cm\epsfbox{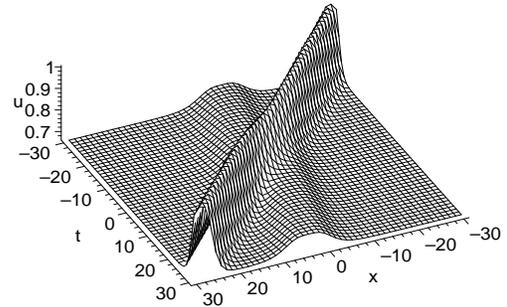}
\caption{Elastic interactions between two kink-antikink molecules of the KK equation described by \eqref{kk4s} with the parameter selections \eqref{kk4s1}.}\label{fig8}
\end{figure}

In summary, soliton molecules can be found in not only the optical systems \cite{Sci2017,LiuXM2018} but also in fluid systems which may be applied for the oceanic surface and internal waves \cite{GPP,KRI}, plasma waves \cite{Plasma}, electromagnetic waves in discrete transmission lines \cite{electro} and so on. The fluid model \eqref{5th} is integrable only in three special cases, the fifth order KdV equation, the SK equation and the KK equation. However, everyone of three integrable models can be used to approximately solve the original nonintegrable fluid system \eqref{5th} up to the same order of $\epsilon$. Though the integrable systems are well known in literature, the soliton molecules and asymmetric solitons of these models have not yet been found before. The soliton molecules for the fifth order KdV and SK systems are quite similar, however, the soliton molecules of the KK equation are very different to those of the fifth order KdV and SK cases. For the SK and fifth order KdV systems, two solitons in the molecule possess different amplitudes and widths while two solitons in the soliton molecule of the KK system possess completely same properties except for the positions. The kink-antikink molecules only exist for the KK equation. Soliton molecules can be considered as some types of special soliton resonance solutions. Soliton molecules are stable under the meaning that the interactions among soliton molecules are elastic. When two solitons in the molecule is closed enough, the molecule looks like one peak soliton which is asymmetric for the SK and fifth order KdV cases and symmetric for the KK equation. Both the soliton molecules and the asymmetric solitons obtained in this letter may be observed in the fluid systems such as the atmospheric and oceanic dynamics, the stratified fluids, the magnetic fluids, the quantum fluids, the fluids of light, the atomic and molecular gases, the degenerate electron fluids, the superfluid
$^3$He \cite{LFH,HHC1979,PRL2005a,PRL2011,RMP2013, RMP2017,RMP2011,RMP1987} and in other physical systems where the model equation \eqref{5th} is approximately valid.

\section*{Acknowledgement}
The work was sponsored by the National Natural Science Foundations of China (Nos. 11435005,11975131) and K. C. Wong Magna Fund in Ningbo University.

\end{document}